\documentclass[aps, prl, reprint, longbibliography, superscriptaddress]{revtex4-2}

 \usepackage{graphicx}

 \usepackage{indentfirst}
\usepackage{placeins}

 \graphicspath{{Figures/}} 

\usepackage{etoolbox} 

 \usepackage{mathrsfs}
 \usepackage{amsfonts}
 \usepackage{amsmath}
 \usepackage{amssymb}
 \usepackage{bm}

\usepackage[version=4]{mhchem}
\usepackage{physics}
\usepackage[separate-uncertainty = true,multi-part-units=single]{siunitx}
\usepackage{natbib}
\usepackage[backref=none,bookmarksnumbered=true,bookmarks=true,bookmarksopen=true,colorlinks=true,citecolor=blue,linkcolor=blue,anchorcolor=green,urlcolor=blue,unicode=false]{hyperref}

\usepackage{ulem}[normalem] 
\def\gnr{2B-575-aGNR}
\def\gnrs{2B-575-aGNRs}
\def\pgnr{2B-575$^*$-aGNR}
\def\pgnrs{2B-575$^*$-aGNRs}

\newcommand*\didv{\mathrm{d} I/\mathrm{d} V}
\newcommand*\diidv{\mathrm{d}^2 I/{\mathrm{d} V}^2}


\begin{document}
\title{Addressing electron spins embedded in metallic graphene nanoribbons}

  \author{Niklas Friedrich}
         \affiliation{CIC nanoGUNE-BRTA, 20018 Donostia-San Sebasti\'an, Spain}

  \author{Rodrigo E. Mench\'on}
        \affiliation{Donostia International Physics Center (DIPC), 20018 Donostia-San Sebasti\'an, Spain}
        
  \author{Iago Pozo}
        \affiliation{CiQUS, Centro Singular de Investigaci\'on en Qu\'{\i}mica Biol\'oxica e Materiais Moleculares, 15705 Santiago de Compostela, Spain}

  \author{Jeremy Hieulle}
         \affiliation{CIC nanoGUNE-BRTA, 20018 Donostia-San Sebasti\'an, Spain}
  
  \author{Alessio Vegliante}
         \affiliation{CIC nanoGUNE-BRTA, 20018 Donostia-San Sebasti\'an, Spain}

  \author{Jingcheng Li}
         \affiliation{CIC nanoGUNE-BRTA, 20018 Donostia-San Sebasti\'an, Spain}

  \author{Daniel S\'anchez-Portal} 
        \affiliation{Donostia International Physics Center (DIPC), 20018 Donostia-San Sebasti\'an, Spain}
        \affiliation{Centro de F\'{\i}sica de Materiales CSIC-UPV/EHU, 20018 Donostia-San Sebasti\'an, Spain}
  
  \author{Diego Pe\~na} \email{diego.pena@usc.es}
        \affiliation{CiQUS, Centro Singular de Investigaci\'on en Qu\'{\i}mica Biol\'oxica e Materiais Moleculares, 15705 Santiago de Compostela, Spain}
        
  \author{Aran Garcia-Lekue} \email{wmbgalea@ehu.eus}
        \affiliation{Donostia International Physics Center (DIPC), 20018 Donostia-San Sebasti\'an, Spain}
        \affiliation{Ikerbasque, Basque Foundation for Science, 48013 Bilbao, Spain}

  \author{Jos\'e Ignacio Pascual} \email{ji.pascual@nanogune.eu}
        \affiliation{CIC nanoGUNE-BRTA, 20018 Donostia-San Sebasti\'an, Spain}
        \affiliation{Ikerbasque, Basque Foundation for Science, 48013 Bilbao, Spain}

\date{\today}

\newpage

\begin{abstract}
Spin-hosting graphene nanostructures are promising metal-free systems for elementary quantum spintronic devices. Conventionally, spins are protected from quenching by electronic bandgaps, which also hinder electronic access to their quantum state. Here, we present a narrow graphene nanoribbon substitutionally doped with boron heteroatoms that combines a metallic character with the presence of localized spin 1/2 states in its interior. The ribbon was fabricated by on-surface synthesis on a Au(111) substrate. Transport measurements through ribbons suspended between the tip and the sample of a scanning tunnelling microscope revealed their ballistic behavior, characteristic of metallic nanowires. Conductance spectra show fingerprints of localized spin states in form of Kondo resonances and inelastic tunnelling excitations. Density functional theory rationalizes the metallic character of the graphene nanoribbon due to the partial depopulation of the valence band induced by the boron atoms. The transferred charge builds localized magnetic moments around the boron atoms. The orthogonal symmetry of the spin-hosting state's and the valence band's wavefunctions protects them from mixing, maintaining the spin states localized. The combination of ballistic transport and spin localization into a single graphene nanoribbon offers the perspective of electronically addressing and controlling carbon spins in real device architectures.
\end{abstract}

\maketitle

\section{}
Graphene nanoribbons (GNRs) are narrow stripes of graphene a few nanometer wide. In spite of graphene being inherently a semimetallic material, electronic correlations and confinement of their electrons into one dimension generally result in gapped band structures \citep{Son2006, Yang2007}. Since the bandgap depends on the GNR's orientation, edge and width \citep{Yang2007, Yazyev2011, chen2015molecular, Talirz2019, Li2021}, precise control of their semiconducting character can be achieved by on-surface synthesis methods \citep{Cai2010, Rizzo2018, Groning2018, clair2019controlling, Li2021}.

In the last years it has been proposed that GNRs can also host localized spin states at specific positions of their carbon lattice, turning them into potential candidates for metal-free spintronic devices \citep{Wang2008, Wang2009, Wu2018, Cao2017}. 
Spin states have been found in GNRs and in graphene nanoflakes, mostly localized around zig-zag edges and various types of defects \citep{Li2019NatCom, Mishra2020NatNano, Mishra2020JACS, Lawrence2020, Mishra2020Angew, Cirera2020, Li2020PRL, sanchez2020diradical, Friedrich2020, Zheng2020, sanchez2020unravelling, Mishra2021NatChem, turco2021surface, Mishra2021Nature, Hieulle2021, Wang2021, biswas2022synthesis, Wang2022, Karan2022, Song2022}.
Two-terminal electronic transport measurements using a scanning tunnelling microscope (STM) have demonstrated that spins can be addressed by electrons tunnelling through the GNR's bandgap \citep{Li2019NL, Friedrich2020}.
Although a bandgap favors spin localization, it restricts low-energy electron movement to distances of a few angstroms. This limits the integration of spin-hosting GNRs into spintronic devices. Ballistic transport through metallic GNRs \citep{Tzalenchuk2010, Aprojanz2018, Karakachian2020} would ease the implementation by facilitating the read out of the embedded spins.

Here, we report on the detection of localized spins in metallic GNRs realized by substitutionally doping a narrow-bandgap GNR with boron atoms in its interior. The boron heteroatoms turn the ribbon metallic and, at the same time, acquire a net magnetic moment. Density functional theory (DFT) calculations reveal that the spin is protected from the partially filled valence band (VB) by the different symmetry of the VB and the boron bands. Pentagonal defects, as those observed in the experiment, break the structural symmetry and open small hybridization gaps in the VB close to the Fermi level. We combined two-terminal transport experiments with differential conductance ($\didv$) spectroscopy to probe the electronic and magnetic properties of individual GNRs. Ballistic transport was stable over distances of several nanometers. The presence of a Kondo resonance proves access to the spin at a transport length of more than $\SI{3}{nm}$.

\section{Results and Discussion}
\begin{figure*}
	\includegraphics[width=2\columnwidth]{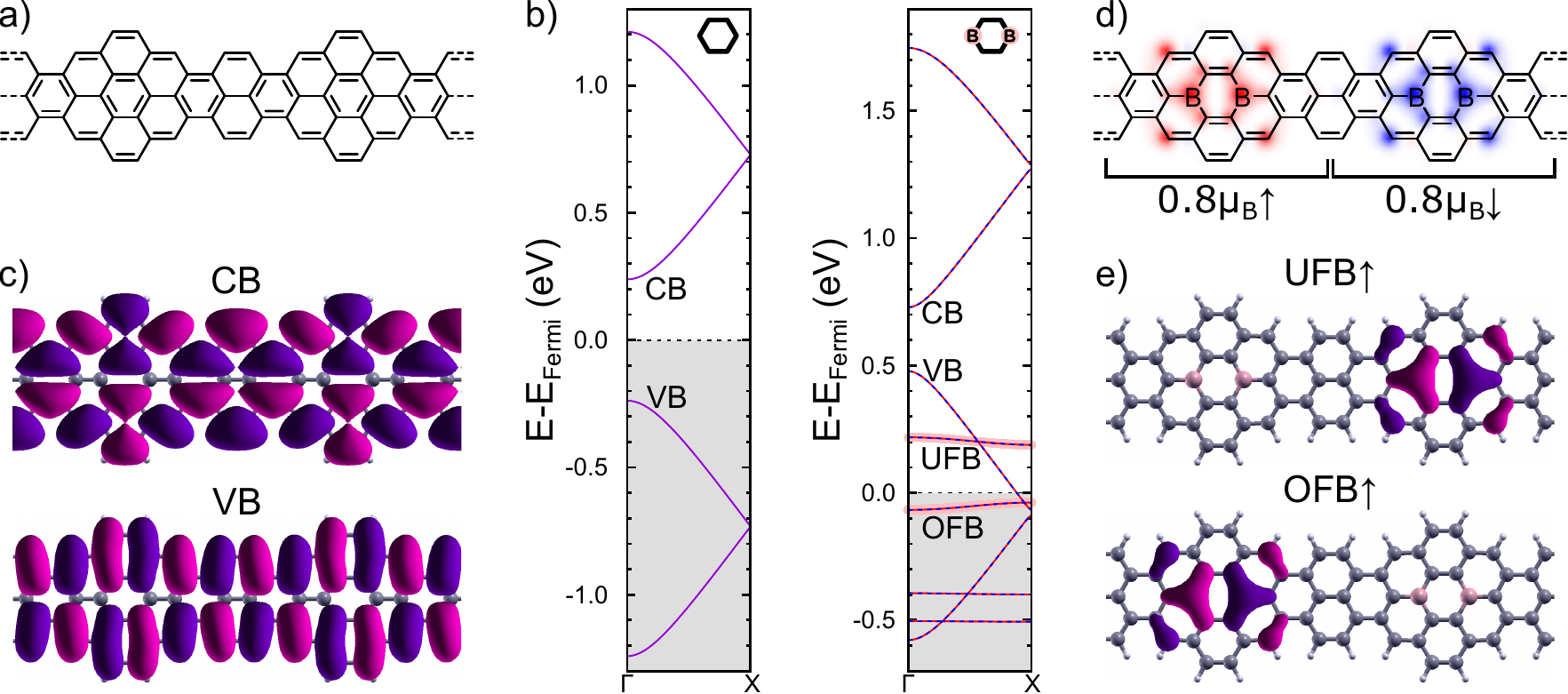}
	\caption{\label{Fig1}
	(a) Lewis structure of the proposed 575-aGNR without boron doping.
    (b) Spin-polarized DFT calculated band structure of the 575-aGNR and the \gnr\ using a doubled supercell like shown in panel (a). Boron character of the bands is indicated by a pink shadow.
    (c) DFT calculated wavefunctions at $\Gamma$ of the CB and VB of the 575-aGNR.
    (d) Lewis structure of the \gnr\ shown on top of a color map representing the calculated spin-polarization density.
	(e) DFT calculated wavefunctions at $\Gamma$ of the spin-up unoccupied (UFB) and occupied (OFB) boron flat band of the \gnr.
	}
\end{figure*}

\textbf{Simulations of the electronic structure of a metallic graphene nanoribbon.} The atomic structure of the investigated ribbon is derived from that of a 7-atom wide armchair GNR (7aGNR) with substitutional boron doping at periodic intervals \citep{Cloke2015, Kawai2015, Carbonell-Sanroma2017nanoletters, Kawai2018, Carbonell-Sanroma2018, Senkovskiy2018, Zhang2022}. Here, we modified the edge structure and width of the undoped 7aGNR by periodically alternating five and seven carbon atom wide segments (575-aGNR, Figure~\ref{Fig1}a).
DFT calculations of the electronic band structure, shown in Figure~\ref{Fig1}b, predict that the undoped GNR has a small bandgap, with no spin polarization. Interestingly, the wavefunction of both valence and conduction bands are antisymmetric with respect to the central axis of the ribbon (Figure 1c), in contrast with the symmetric character of frontier bands in the related 7aGNR \citep{Carbonell-Sanroma2018}. This change in the bands' symmetry turns out to be crucial to understand the effect of boron substitution inside the GNR. 

Substituting the two central carbon atoms in the wider segments with boron atoms (as shown in Figure~\ref{Fig1}d) creates two boron-rich flat bands. These bands originate from pure boron orbitals and have no topological character, unlike the flat bands of 2B-7aGNRs \citep{Carbonell-Sanroma2018, Friedrich2020}. The band structure of the \gnr, also shown in Figure~\ref{Fig1}b, reveals a significant charge transfer from its VB to the boron bands, resulting in an occupied flat band (OFB), hosting approximately 2 electrons, and an unoccupied flat band (UFB). The VB is partially depopulated and becomes metallic.

The boron flat bands and VB cross without opening a gap, as shown in Figure~\ref{Fig1}b, revealing a negligible mixing. This is a consequence of the different symmetries of their wavefunctions (Figure~\ref{Fig1}c, e). The boron flat bands, localized around the boron atoms, are symmetric with respect to the central ribbon axis, while the VB is anti-symmetric. The orthogonality between VB and boron flat bands allows electrons in the VB to propagate unperturbed along the ribbon \citep{Carbonell-Sanroma2017nanoletters}, resulting in a metallic band and maintaining the boron states localized around the diboron impurities. DFT finds a magnetic moment of $0.8\mu_\text{B}$, close to spin $S=1/2$, associated with the OFB that is localized around each 2B-unit. Spin moments in adjacent 2B-units tend to anti-align, as shown in Figure~\ref{Fig1}d, so the periodic system shows no net spin-polarization.

\begin{figure}[hbt]
	\includegraphics[width=1\columnwidth]{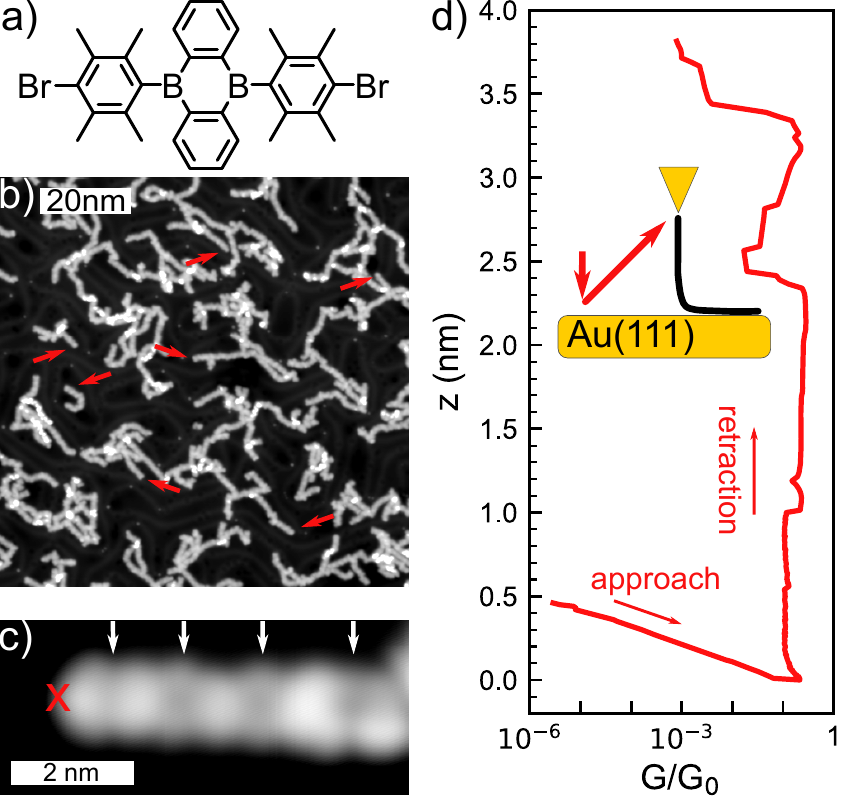}
	\caption{\label{Fig2}
	(a) Molecular precursor used for synthesizing the \gnrs .
    (b) STM topography image ($V = \SI{1}{V}$, $I=\SI{30}{pA}$). Straight \gnrs\ are indicated by red arrows.
	(c) STM topography image ($V = \SI{-300}{mV}$, $I=\SI{30}{pA}$) of a four precursor unit long \gnr. The positions of boron doping are indicated by white arrows. The red cross indicates the position from where the GNR is lifted for the transport experiment.
    (d) $G(z, V = \SI{10}{mV})$ for the GNR presented in (c). The conductance is independent of $z$ up to $z\approx \SI{2.2}{nm}$. The inset is a schematic drawing of the experimental setup.
    }
\end{figure}

\textbf{On-surface synthesis of \gnrs .} Based on the intriguing properties predicted by DFT calculations, we decided to explore the on-surface synthesis and characterization of this boron-doped GNR. A retrosynthetic analysis identified the compound shown in Figure~\ref{Fig2}a as the ideal molecular precursor, which might lead to the formation of \gnrs\ by sequential Ullmann coupling and cyclodehydrogenation reactions on a Au(111) substrate. The molecular precursor was obtained by solution chemistry in one step from easily available starting materials (see Supporting Information (SI) for details) and sublimated \emph{in situ} on Au(111). Polymerization occurs at $\SI{250}{\celsius}$, a higher temperature than for other systems \citep{Cai2010, Kawai2015}, and close to the onset of cyclodehydrogenation of the polymer. The presence of the precursor's bulky methyl groups increases the energy barrier for the formation of metal organic complexes \citep{Fritton2018}, which have been shown to facilitate the on-surface Ullmann coupling \citep{Barton2017}. As a consequence, the Ullmann coupling requires a higher temperature for activation. We annealed the sample to $\SI{300}{\celsius}$ to achieve a high amount of planar ribbons. The resulting structures were mostly curved and interlinked ribbons, as seen in Figure~\ref{Fig2}b, with a few short and straight segments (red arrows). Figure~\ref{Fig2}c shows an STM image of a single straight \gnr\ segment, where four boron doping sites can be identified as wider and darker segments of the ribbon.

\textbf{Two-terminal electronic transport measurements.} We studied the electronic transport through a GNR suspended between the tip and sample of an STM \citep{Lafferentz2009, Koch2012, Jacobse2018, Li2019NL, Friedrich2020, Lawrence2020}. To reach this two-terminal configuration, we positioned the tip above the apex of the \gnr\ (red cross in Figure~\ref{Fig2}c) and approached the tip towards the substrate until a sudden increase in the current indicated the formation of a bond between tip and ribbon. Then, we retracted the tip following a leaned trajectory along the backbone of the ribbon. This procedure lifts the \gnr\ partially from the Au(111) and electronically decouples the free-standing segment from the metal.

Electronic transport measurements through the lifted ribbons confirm that they behave as ballistic conductors. As shown in Figure~\ref{Fig2}d, the linear conductance $G(z)$ of the ribbon remains constant for several nanometers while the tip is retracted. The constant conductance contrasts with the exponentially decaying conductance found for semiconducting ribbons \citep{Lafferentz2009, Chong2018, Li2019NL, Friedrich2020}. In a ballistic conductor, the electron transmission $\mathcal{T}$ remains constant as a function of its length, and the conductance per channel amounts to $\mathcal{T} G_0$, where $G_0=e^2/\pi\hbar=\SI{77.5}{\mu S}$ is the conductance quantum. In the results shown in Figure~\ref{Fig2}, we observe high conductance values around $\sim 0.2 G_0$ remaining constant for more than $\SI{2}{nm}$ of GNR elevation. The electron transmission smaller than $G_0$ is probably caused by the finite contact resistance between tip and GNR \citep{Beebe2002, Venkataraman2006}. At some points we find small variations of the conductance around $0.1 G$, which are consistent with atomic-scale rearrangements of the GNR-electrode contacts when additional borylated units detach from the surface \citep{Lafferentz2009, Reecht2014}. The ballistic electron transport found here reflects the existence of scattering free transmission channels in free-standing \gnrs, in agreement with the boron induced metallic character revealed by our DFT calculations in Figure~\ref{Fig1}.

\begin{figure}
	\includegraphics[width=1\columnwidth]{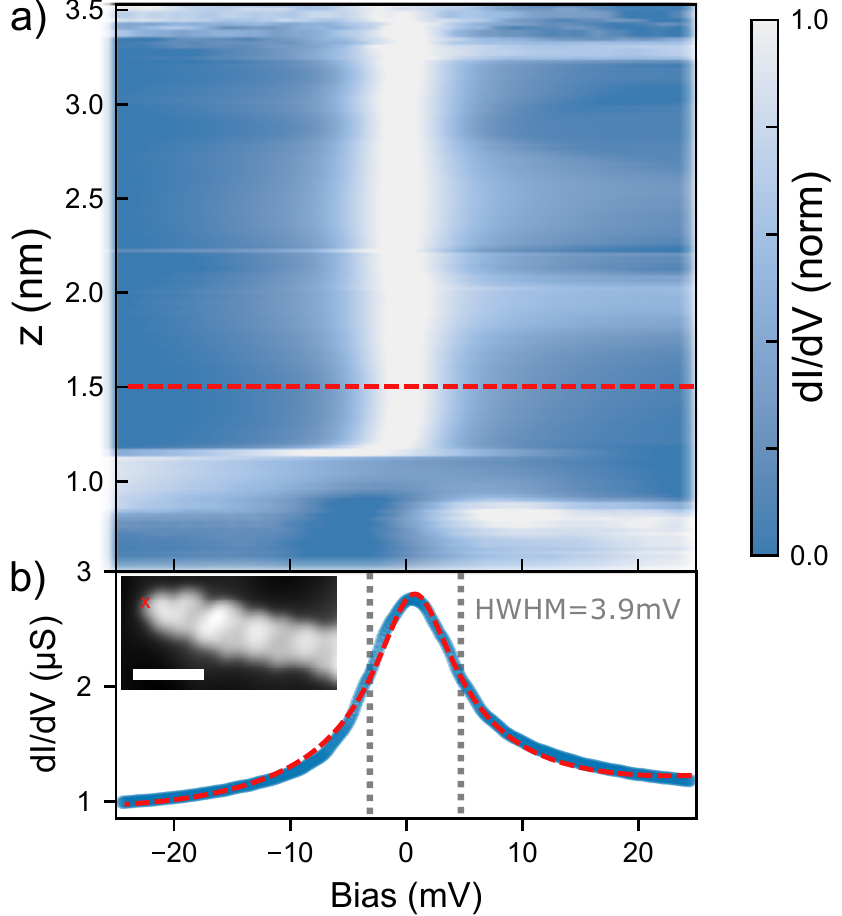}
	\caption{\label{Fig3}
	(a) Normalized $\didv (V,z)$-map (see methods for details on normalization procedure). A zero-bias resonance appears in the spectra for $z>\SI{1.2}{nm}$.
	(b) One example $\didv$-spectrum (blue) fitted with a Frota function \citep{Frota1992} (red dashed line). The spectrum was taken at $z=\SI{1.5}{nm}$. Inset: STM topography image ($V = \SI{-300}{mV}$, $I=\SI{30}{pA}$, scale bar is $\SI{2}{nm}$). The red cross indicates the position from where the GNR is lifted.
    }
\end{figure}

\textbf{Observation of the Kondo effect in ballistic ribbons.}
Figure~\ref{Fig3}a shows a $\didv (V,z) $ spectral map obtained by measuring $\didv$ spectra during the lift of a \gnr\ (inset Figure~\ref{Fig3}b). A narrow zero-bias resonance ($\text{HWHM}=\SI{3.2}{mV}$) appears suddenly at $z=\SI{1.2}{nm}$ and prevails up to $\Delta z\geq \SI{3.5}{nm}$ during the lifting procedure. We interpret this resonance as a manifestation of the Kondo effect in the electronic transport through the ribbon \citep{Kondo1964}. The Kondo resonance is the fingerprint of a spin state weakly coupled to an electron bath \citep{Anderson1966, Roch2009, Ternes2009, Li2019NatCom, Lawrence2020, Li2020PRL, Friedrich2020}. Here, it is observed for more than $\SI{2}{nm}$ during the GNR elevation, hinting that the Kondo screening is not simply mediated by electrons at the surface \citep{Friedrich2020}. We suggest that the metallic band of the ribbon is responsible for the screening of the localized magnetic moments. 

\begin{figure*}[ht!]
	\includegraphics[width=2\columnwidth]{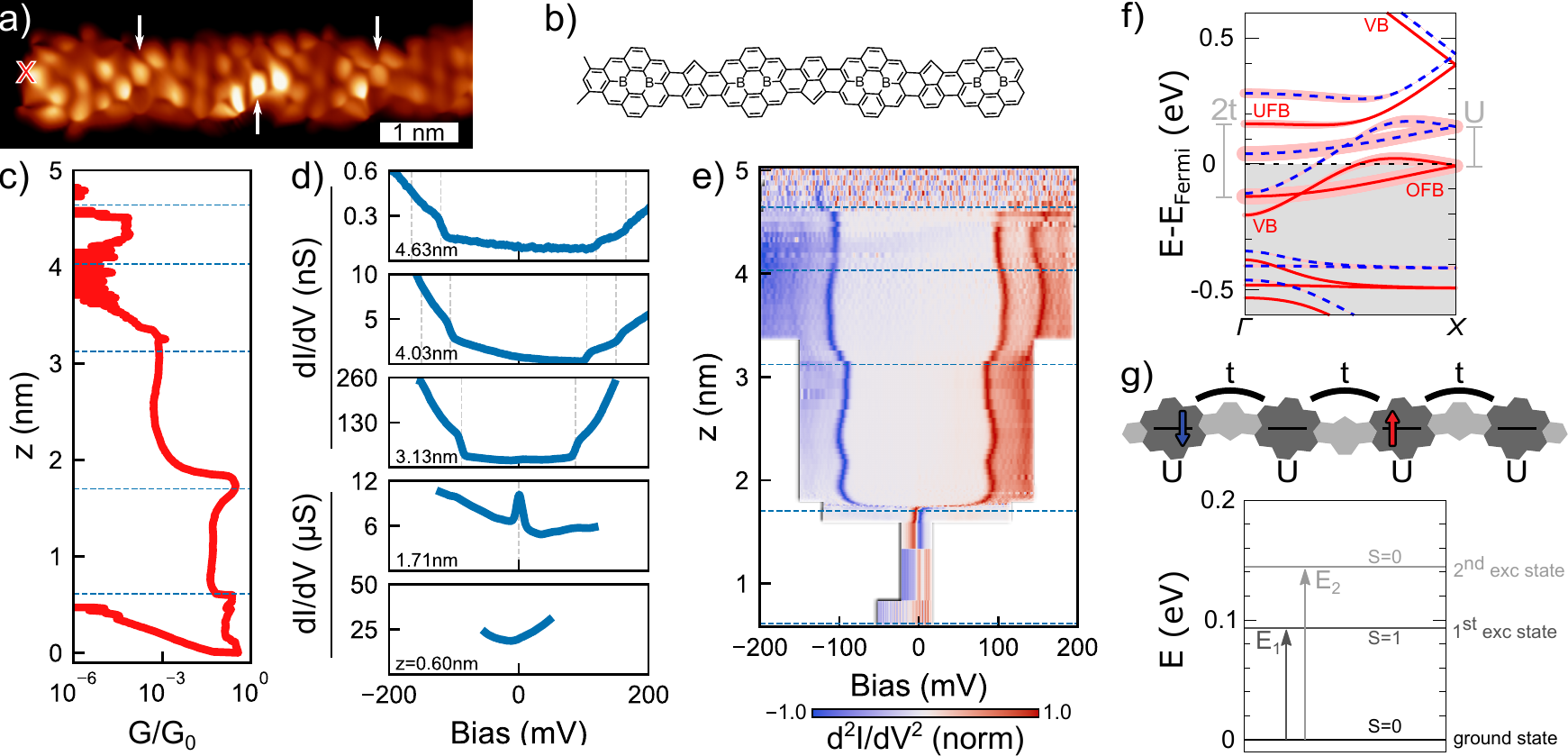}
	\caption{\label{Fig4}
	(a) Bond-resolved constant height current image ($V = \SI{5}{mV}$). The 2B-units alter the contrast due to buckling of the ribbon. Pentagons are indicated by white arrows.
	The red cross indicates the position from where the ribbon is lifted.
	(b) Lewis structure of the \pgnr\ in (a).
	(c) Linear conductance $G(z, V = \SI{10}{mV})$ obtained while lifting the ribbon presented in (a).
	Some ballistic behavior is retained.
	(d) $\didv$-spectra at selected $z$. The values are indicated and correspond to the blue, dotted lines in (c) and (e). A single Kondo-resonance at $z=\SI{1.71}{nm}$ indicates the presence of a spin $S=1/2$. For larger $z$ inelastic excitations dominates the spectra.
	(e) Normalized $\diidv (V,z)$-map obtained by numerical differentiation (see methods for details on normalization procedure).
	(f) DFT calculated band structure of periodic \pgnr . Notice that in this case the unit cell necessarily contains two 2B-units. Red and blue bands correspond to spin up and down. Boron character of the bands is indicated by a pink shadow. The corresponding band from the \gnr\ is indicated for the spin up band.
	(g) Hubbard-model used to calculate spin-excitations of the finite sized system and its energy spectrum for two electrons that are delocalized across four hopping sites. $U=\SI{155.9}{meV}$, $t=\SI{133.9}{meV}$, obtained from (f).
}
\end{figure*}

In the transport experiments presented in Figure~\ref{Fig2} and \ref{Fig3} we found electronic and magnetic fingerprints that are consistent with our DFT calculations of the \gnr . All the 19 ribbons explored in this manner show similar results, in every
case reproducing segments of constant conductance associated to ballistic transport, and Kondo resonances. However, many ribbons also showed a step-wise decrease in their conductance plots at some elongations, which we attribute to the presence of defects in their structure.

\textbf{Role of atomic defects on the nanoribbons.} To detect the presence of atomic-scale defects in \gnrs, we measured constant height current images using a CO-functionalized tip \citep{Gross2009, Kichin2011}. The image of a nanoribbon composed of 4 molecular units (\emph{i.e.} 4 boron dimers) is shown in Figure~\ref{Fig4}a. It is consistent with a ribbon containing a sequence of pentagonal rings in its carbon backbone at the position of Ullmann coupling (white arrows in Figure~\ref{Fig4}a). The extracted Lewis structure of the ribbon (referred to as \pgnr ) is shown in Figure~\ref{Fig4}b. Pentagonal rings are known to appear when methyl groups of the precursor detach during the polymerization reaction \citep{Mishra2020JACS, Zheng2020, Lu2021}. This type of defect at the linking position is the most common structure we find in an analysis of 16 ribbons.

To unravel the effect of the atomic defects on the electronic transport of \gnrs, we performed two-terminal transport measurements on the ribbon shown in Figure~\ref{Fig4} in a suspended geometry. 
As depicted in Figure~\ref{Fig4}c, the conductance $G(z)$ decreases step-wise with increasing $z$. The conductance steps are spaced by $\Delta z \sim \SI{1.4}{nm}$, matching with the distance between two diboron sites. This suggests that they appear when a new precursor unit is inserted in the free-standing part of the ribbon \citep{Reecht2014}.

Between conductance steps, constant conductance plateaus unveil that some ballistic behavior is retained. However, now, we found three qualitatively different regimes, depicted in Figure~\ref{Fig4}d. First, we observed a metallic like behavior, with flat $\didv \sim 0.3 G_0$ signal persisting until $z=\SI{0.6}{nm}$, where the first conductance steps appears. Upon further tip retraction, a zero-bias resonance similar to the one shown in Figure~\ref{Fig3} appears in the spectra. Again, this indicates that a localized spin appears in the free-standing segment of the ribbon. This Kondo feature disappears at $z=\SI{1.75}{nm}$, coinciding with a second step in the linear conductance plot of Figure~\ref{Fig4}c. Above this $z$ value, spectra exhibit two bias-symmetric $\didv$ steps, characteristic of inelastic electron tunnelling (IET) excitations.

To follow the IET spectral evolution during the retraction, we show in Figure~\ref{Fig4}e a normalized $\diidv (V,z)$-map. We observe that at $z=\SI{1.75}{nm}$ the Kondo resonance splits gradually in $\leq \SI{1}{\angstrom}$ and converts into IET steps (see SI Figure~4).
Above this value, the steps are observed for more than $\SI{3}{nm}$ retraction with small variations of their excitation energy. A fainter $\didv$-step, at approximately $\SI{45}{mV}$ larger bias voltage, can also be observed in the spectra above $\SI{3}{nm}$. The continuous evolution from Kondo to IET excitations suggests that a complex spin texture exists in the \pgnr .

DFT calculations for periodic \pgnrs\ revealed that the presence of the pentagonal rings in the ribbon has two important implications. First, they break the structural symmetry of the GNR, mixing the wavefunctions of boron flat bands and VB. Now, the band structure in Figure~\ref{Fig4}f shows avoided crossings of both UFB and OFB with the VB, characteristic of a small hybridization. Second, the removal of a carbon atom from the ribbon effectively injects another hole and lowers the occupation of both the VB and the OFB. Since the VB still crosses the Fermi level, the ribbon preserves its metallic character. DFT pictures the depopulation of the OFB as a mean delocalization of an electron over several 2B-units and a smaller net magnetic moment associated with each diboron unit ($\sim 0.5 \mu_B/\text{2B}$).

To interpret the experimental IET signal, we explored different magnetic states obtained by DFT simulations of finite ribbons, like the one presented in Figure~\ref{Fig4}g (see SI Figure~7). DFT results indicate that the system cannot be simply treated as Heisenberg-like Hamiltonian due to the electron delocalization. In fact, DFT significantly underestimates the excitation energies as compared to the measured IET spectra and did not fully capture the relevant physical mechanisms behind the inelastic steps.

Given that the hybridization between the VB and the boron flat bands is small (notice in Figure~\ref{Fig4}f the 1 to 10 ratio between the size of the hybridization gaps and the VB band width), a valid approximation is treating VB and OFB as two different subsystems, 
disregarding excitations that imply charge-transfer between them. 
The observed spectral steps can then be attributed to inelastic excitations in the OFB/UFB subsystem induced by conduction electrons propagating through the VB. 
To describe the excitation spectrum of the boron flat bands, we use a Hubbard model with parameters $t$ and $U$ obtained from the DFT band structure in Figure~\ref{Fig4}f. This simple model can be exactly solved and approximately accounts for electron correlations in the excitation spectrum. Based on the OFB's occupation observed in DFT calculations (see Figure~\ref{Fig4}f), we consider the probable case of two electrons distributed over four electron sites.

Exact diagonalization of the Hubbard Hamiltonian leads to the energy spectrum presented in Figure~\ref{Fig4}g. The ground state is a singlet combination of the two spins. The first excited state is a triplet state located at $E_1 = \SI{94}{meV}$, which matches reasonably well with the energy of the first excitation step in our experiments. Furthermore, the model also finds an excited singlet state at $E_2 = \SI{145}{meV}$, in strong coincidence with the second spectral IET step. The observed agreement between the calculated and experimentally observed excitation energies indicates the IET signals can be due to electron-hole pair excitations of a partially populated flat band, a similar excitation process to the one recently observed on small molecules on insulating layers \citep{Fatayer2021}. This interpretation is supported by the fact that the excitation energies in the model are dominated by the hopping parameter between two 2B-sites $t$ and are stable against changes of $U$ (SI Figure~5) or electron occupation (SI Figure~6). This explains why the step values are relatively independent from the length of the free-standing segment. The model unveils a probable source of scattering phenomena in the electronic transport, which, also, can be responsible for the step-wise decrease of the linear conductance found when a new diboron unit is lifted from the surface (Figure~\ref{Fig4}c).

\section{Conclusions}
We have presented a graphene nanoribbon that combines a metallic transport band with localized spins inside, and described experimental fingerprints of both the metallic character and the spin-polarized states. The ribbon is fabricated by substitutionally doping the narrow-bandgap 575-aGNR with diboron units. DFT simulations unveiled that the dispersive VB becomes partially depopulated by donating electrons to flat bands formed by the diboron units. As a result, single spins emerge localized at every diboron unit, protected from the VB by their different wavefunction symmetry. Two-terminal transport experiments through free-standing GNRs placed between tip and sample of an STM demonstrated ballistic electron transport, transmitting electrons with a constant value of $0.2 G_0$ for a free-standing segment length of a few nanometers. Simultaneously, the differential conductance spectroscopy in the transport configuration unraveled the spin localization by revealing a Kondo resonance that was stable during the elevation of the GNR from the surface. 

In addition to the ideal case, we found that atomic defects such as pentagonal rings in the structure, frequently found in our experiments, enable a small wavefunction mixing between VB and the boron flat bands and partially depopulate both bands. The effect of these pentagonal defects in the transport is drastic, because it enables a finite interaction between adjacent diboron units and delocalizes the electrons along the partially occupied flat band. We found that the ballistic character of the ribbon partly survives in the presence of pentagonal defects, but a new inelastic excitations appears in the spectra, accompanying a step-wise decrease of the linear conductance with ribbon elevation.
Through a simple Hubbard model, parameterized with results from the DFT simulations, we found that the inelastic spectral features can be attributed to excitation of the many-body states of the partially depopulated flat band. These results thus demonstrate that the \gnr\ represent a unique molecular system to import flat band phenomena into one-dimensional graphene nanoribbons, envisioning the study of the underlying electron transport phenomena present in these correlated systems.

\section{Experimental and Methods}
\textbf{Sample preparation.} A Au(111) single crystal was prepared by \ce{Ne^+} sputtering and successive annealing at $T=\SI{450}{\celsius}$ under ultra-high vacuum conditions. The precursor molecules were sublimated \emph{in situ} from a Knudsen cell at a temperature of $\SI{220}{\celsius}$. Afterwards the gold was annealed to $T=\SI{180}{\celsius}$ for $10$ minutes and flashed to $T=\SI{300}{\celsius}$ for one minute. The samples were analyzed in a custom-made low-temperature STM at $\SI{5}{K}$. The Figures presenting experimental data were prepared using WSxM \citep{horcas2007wsxm} and the python matplotlib libary \citep{Hunter2007matplotlib} using perceptual continuous color scales \citep{kovesi2015good}.

\textbf{Differential conductance measurements.} Spectroscopic $\didv$-measurements were performed using an external lock-in amplifier with frequency $f=\SI{867.6}{Hz}$, time constant $\tau = \SI{30}{ms}$ and modulation $V_\text{mod} = \SI{2}{mV}$. The $G(V, z)$-map in Figure~\ref{Fig3}a is normalized for each $z=z_0$ separately, using the formula $G_\text{norm}(V, z_0) = (G(V, z_0) - G_\text{min})/(G_\text{max}-G_\text{min})$, where $G(V, z_0)$ is the differential conductance and $G_\text{max(min)}$ the maximum (minimum) value of $G(V, z_0)$. The $\diidv$-map in Figure~\ref{Fig4}e is normalized equivalently using $G'_\text{norm}(V, z_0) = G'(V, z_0)/|G'_\text{min(max)}|$, where $G'$ is the derivative of the conductance and $G'_\text{min(max)}$ is the minimum (maximum) of $G'$. $G'_\text{min(max)}$ was used for negative (positive) values of $G'(V)$.

\textbf{Density functional theory calculations.} First-principles electronic structure calculations were performed using DFT as implemented in the SIESTA software package\citep{Artacho1999, Soler2002}. The van der Waals density functional by Dion \textit{et al.} \citep{Dion2004} with the modified exchange correlation by Klime{\v{s}}, Bowler and Michaelides\citep{Klimes2009} was used. The valence electrons were described by a double-$\zeta$ plus polarization (DZP) basis set with the orbital radii defined using a $\SI{54}{meV}$ energy shift \citep{Soler2002}, while the core electrons were described using norm-conserving Trouillers-Martins pseudopotentials \citep{Troullier1991}. For integrations in real space\citep{Soler2002} an energy cutoff of $\SI{300}{Ry}$ was used. The smearing of the electronic occupations was defined by an electronic temperature of $\SI{300}{K}$ with a Fermi-Dirac distribution. The self-consistency cycles were stopped when variations on the elements of both the density matrix and the Hamiltonian matrix were smaller than $10^{-4}\SI{}{eV}$. In order to avoid interactions with periodic images from neighboring cells, systems were calculated within a simulation cell where at least $\SI{50}{\angstrom}$ of vacuum space was considered. Variable cell relaxations and geometry optimizations were performed using the conjugate gradient method using a force tolerance equal to $\SI{10}{meV/\angstrom}$ and $\SI{0.2}{GPa}$ as a stress tolerance. For periodic ribbons, a 40 k-point mesh along the GNRs' periodic direction was used.

\section{Acknowledgments}
We gratefully acknowledge financial support from Spanish AEI and the European Regional Development Fund (ERDF) and the Maria de Maeztu Units of Excellence Program and from the European Union (EU) through Horizon 2020 (FET-Open project SPRING Grant. no. 863098).
We thank the Spanish Agencia Estatal de Investigación (PID2019-107338RB-C62, PID2019-110037GB-I00 and PCI2019-111933-2 and Xunta de Galicia (Centro de Investigación de Galicia accreditation 2019–2022, ED431G 2019/03) and Dpto. Educación Gobierno Vasco (Grant No. PIBA-2020-1-0014) and Programa Red Guipuzcoana de Ciencia, Tecnología e Innovación 2021 (Grant Nr. 2021-CIEN-000070-01. Gipuzkoa Next) for financial support.


%

\end{document}


\title{Addressing electron spins embedded in metallic graphene nanoribbons - Supporting Information}

  \author{Niklas Friedrich}
         \affiliation{CIC nanoGUNE-BRTA, 20018 Donostia-San Sebasti\'an, Spain}

  \author{Rodrigo E. Mench\'on}
        \affiliation{Donostia International Physics Center (DIPC), 20018 Donostia-San Sebasti\'an, Spain}
        
  \author{Iago Pozo}
        \affiliation{CiQUS, Centro Singular de Investigaci\'on en Qu\'{\i}mica Biol\'oxica e Materiais Moleculares, 15705 Santiago de Compostela, Spain}

  \author{Jeremy Hieulle}
         \affiliation{CIC nanoGUNE-BRTA, 20018 Donostia-San Sebasti\'an, Spain}
  
  \author{Alessio Vegliante}
         \affiliation{CIC nanoGUNE-BRTA, 20018 Donostia-San Sebasti\'an, Spain}

  \author{Jingcheng Li}
         \affiliation{CIC nanoGUNE-BRTA, 20018 Donostia-San Sebasti\'an, Spain}

  \author{Daniel S\'anchez-Portal} 
        \affiliation{Donostia International Physics Center (DIPC), 20018 Donostia-San Sebasti\'an, Spain}
        \affiliation{Centro de F\'{\i}sica de Materiales CSIC-UPV/EHU, 20018 Donostia-San Sebasti\'an, Spain}
  
  \author{Diego Pe\~na} \email{diego.pena@usc.es}
        \affiliation{CiQUS, Centro Singular de Investigaci\'on en Qu\'{\i}mica Biol\'oxica e Materiais Moleculares, 15705 Santiago de Compostela, Spain}
        
  \author{Aran Garcia-Lekue} \email{wmbgalea@ehu.eus}
        \affiliation{Donostia International Physics Center (DIPC), 20018 Donostia-San Sebasti\'an, Spain}
        \affiliation{Ikerbasque, Basque Foundation for Science, 48013 Bilbao, Spain}

  \author{Jos\'e Ignacio Pascual} \email{ji.pascual@nanogune.eu}
        \affiliation{CIC nanoGUNE-BRTA, 20018 Donostia-San Sebasti\'an, Spain}
        \affiliation{Ikerbasque, Basque Foundation for Science, 48013 Bilbao, Spain}

\date{\today}

\maketitle

\onecolumngrid
\tableofcontents

\clearpage

\section{Synthesis of the molecular precursor}
\FloatBarrier
\subsection{General methods}
Reactions were carried out under argon using oven-dried glassware. \ce{Et2O} and toluene were purified in a MBraun SPS-800 Solvent Purification System. 9,10-dibromo-9,10-dibo- raanthracene (\ce{DBA}) was prepared following a published procedure \citep{bieller2004bitopic} and stored in a glovebox. Other commercial reagents were purchased from ABCR GmbH or Sigma-Aldrich and were used without further purification. TLC was performed on Merck silica gel 60 F254 and chromatograms were visualized with UV light ($254$ and $\SI{365}{nm}$). Column chromatography was performed on Merck silica gel 60 (ASTM 230-400 mesh). \ce{^1H} NMR spectra were recorded at $\SI{300}{MHz}$ (Varian Mercury-300 instrument).

\subsection{Synthesis of precursor molecule}
\begin{figure*}[h!]
    \centering
    \includegraphics{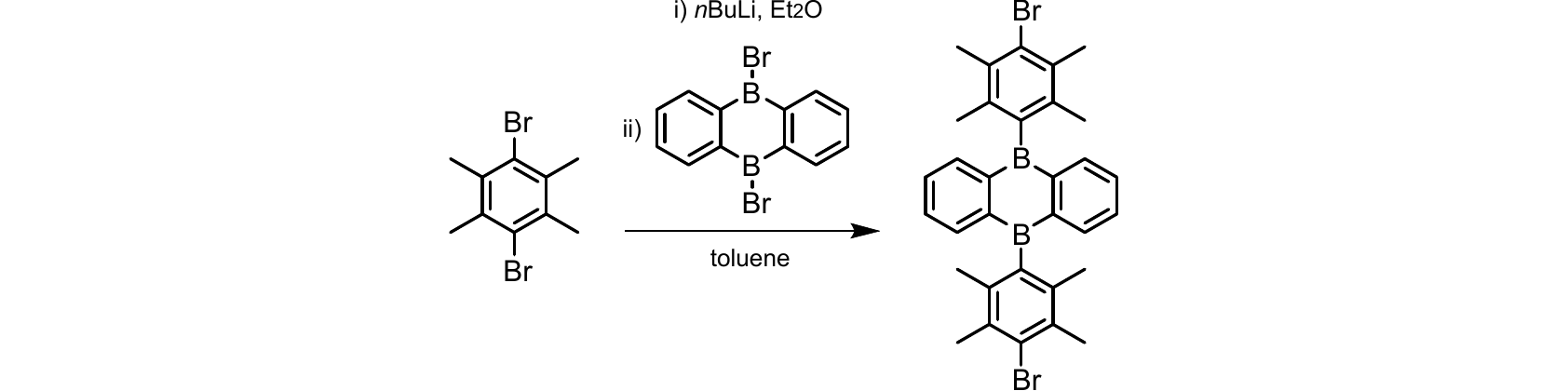}
    \caption{Synthetic route to obtain the boron-doped GNR precursor.}
    \label{fig:SFig00}
\end{figure*}

To a solution of 9,10-dibromodurene ($\SI{174}{mg}$, $\SI{0.60}{mmol}$) in dry \ce{Et2O} ($\SI{0.04}{M}$) at $\SI{-78}{\celsius}$ in a Schlenk tube, \ce{n-BuLi} ($\SI{2.5}{M}$ in hexane, $\SI{0.63}{mmol}$) was dropwise added. Subsequently, the reaction mixture was stirred at $\SI{0}{\celsius}$ for $\SI{20}{min}$ and cooled again to $-\SI{78}{\celsius}$. Then, another solution of 9,10-dibromo-9,10-diboraanthracene (\ce{DBA}, $\SI{100}{mg}$, $\SI{0.30}{mmol}$) in dry toluene ($\SI{0.02}{M}$) was dropwise added at $-\SI{78}{\celsius}$ and allowed to warm up to $\text{rt}\SI{}{\celsius}$ for $\SI{18}{h}$. Then, the solvent was removed under reduced pressure and the crude was purified by column chromatography (\ce{SiO2}, \ce{CHCl3}:hexane 1:1, $\text{Rf} = 0.9$). The solid obtained was washed with hexane and centrifuged affording 5,10-bis(4-bromo-2,3,5,6-tetramethylphenyl)-5,10-dihydroboranthrene as a white solid ($\SI{27}{mg}$, m.p.: $>350$ decomp., $15\%$ yield). \ce{^1H} NMR ($\SI{300}{MHz}$, \ce{CDCl3}): 7.59 (m, 4H), 7.47 (m, 4H), 2.44 (s, 12H), 2.03 (s, 12H) ppm. MS (APCI), m/z: 598 (M+, 100). HRMS (APCI), m/z found: 596.1064 (calc. for \ce{C32H32B2Br2}: 596.1051).
\begin{figure}[h!]
    \centering
     \includegraphics{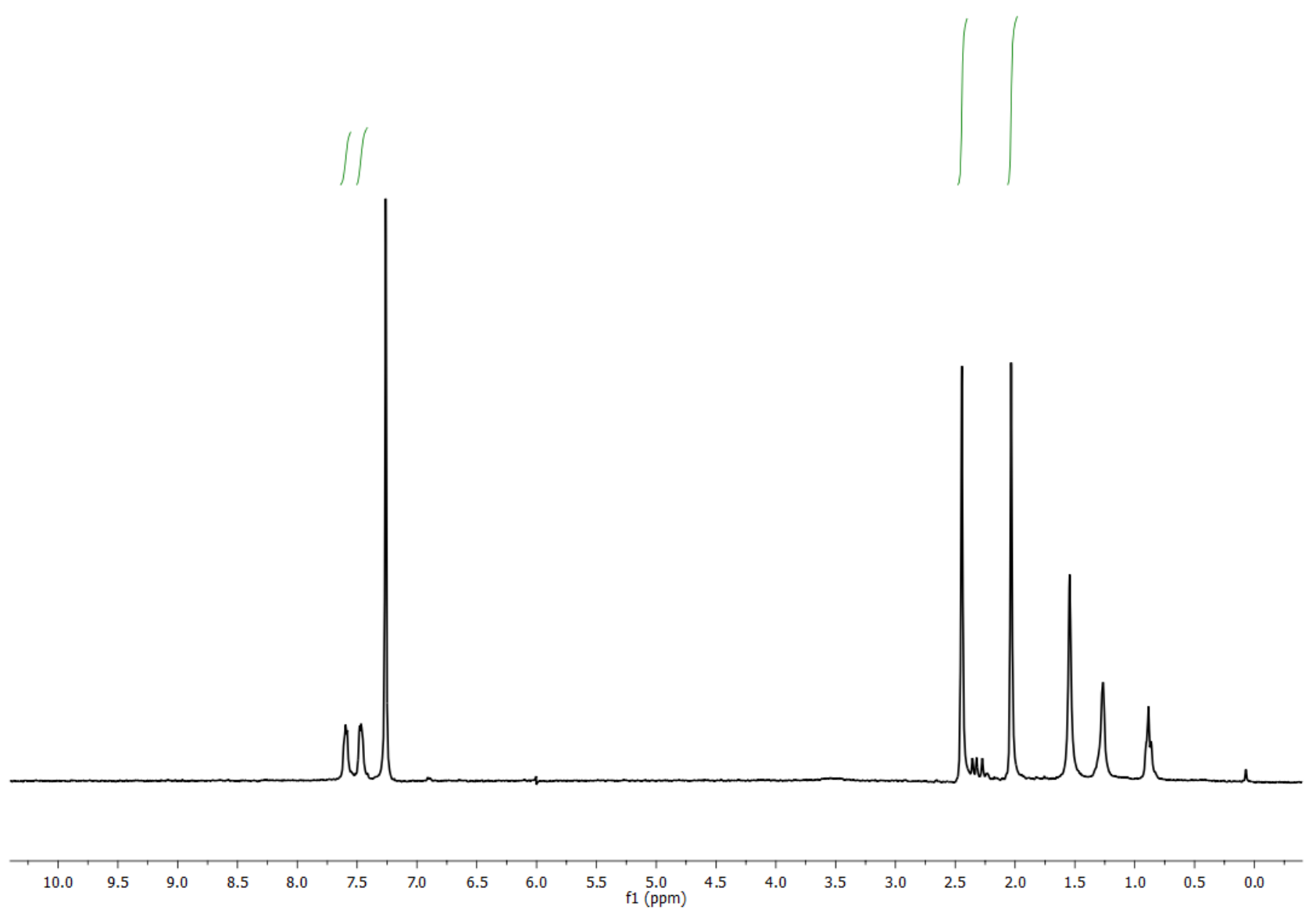}
    \caption{\ce{^1H} NMR of the boron-doped GNR precursor in \ce{CDCl3} at room temperature}
    \label{fig:SFig01}
\end{figure}

\clearpage
\FloatBarrier
\section{Complementary experimental data}
\subsection{Conductance\,-\,Distance Plot}
\FloatBarrier
\begin{figure}[h]
	\includegraphics{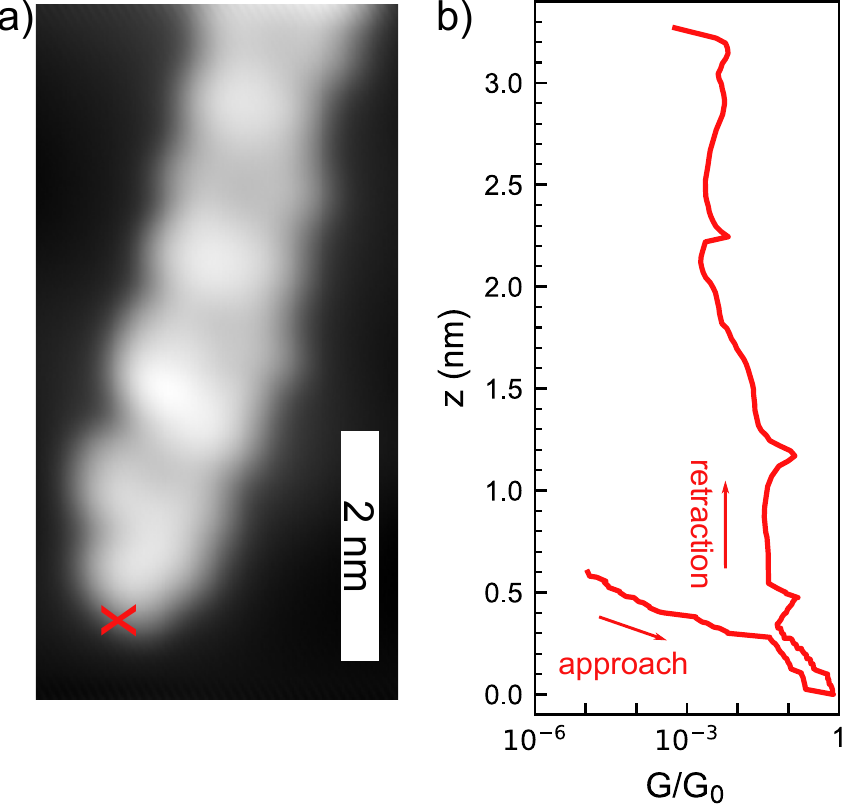}
	\caption{\label{Fig2}
	(a) STM topography image ($V=\SI{-300}{mV}$, $I=\SI{30}{pA}$) of a \gnr. It is the same image as Figure~3b of the main text and only shown for reference.
    (b) $G(z, V = \SI{10}{mV})$ for the GNR presented in (a). The data was recorded simultaneously to the differential conductance map presented in Figure~3a.
    }
\end{figure}
The $G(z)$ retraction curve recorded during lifting the ribbon presented in the main manuscript in Figure~3 initially reaches $G(z)\sim 0.1 G_0$. The conductance lowers with increasing tip sample separation, but does not follow the exponential decay characteristic of semi-conducting ribbons.
\FloatBarrier

\clearpage
\FloatBarrier
\subsection{Transition from Kondo to IET-regime}
\FloatBarrier
\begin{figure}[h]
	\includegraphics{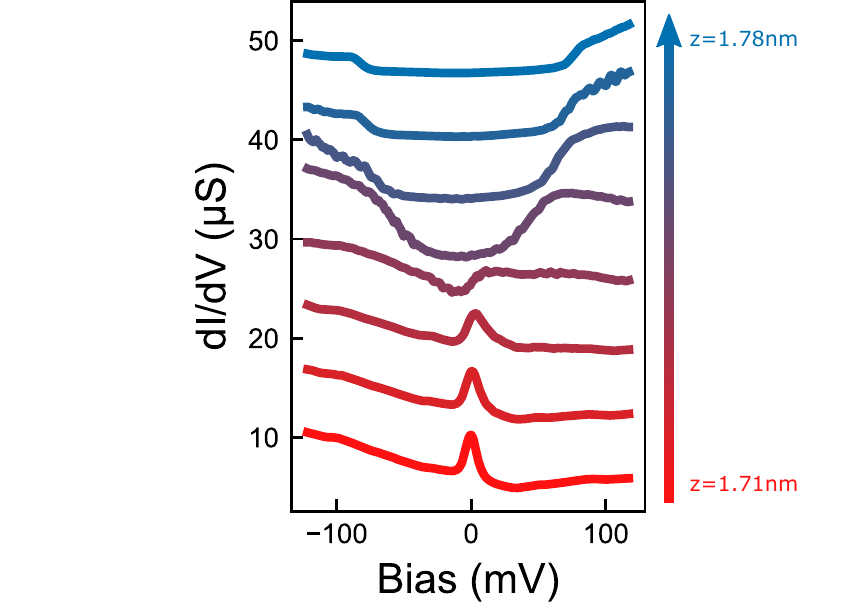}
	\caption{\label{Fig2.5}
	A Stack of differential conductance spectra taken in steps of $\Delta z = \SI{10}{pm}$ shows the continuous transition from the Kondo resonance to a IET excitation. Data corresponding to Figure~4e of the main manuscript. The spectra are offset by $\SI{6.6}{\mu S}$ for clarity.
    }
\end{figure}
The ribbon presented in Figure~4 of the main manuscript undergoes a transition from Kondo regime to an IET regime upon tip retraction starting from $z\sim\SI{1.7}{nm}$. The transition takes place smoothly during few tens of $\SI{}{pm}$. 
\FloatBarrier

\clearpage
\section{The Four Sites Fermi-Hubbard Model}
\FloatBarrier

The finite-size Fermi-Hubbard model with $n$ sites has the following Hamiltonian:
$$\hat{H}= \sum_{j=1}^{n-1} \sum_{\sigma=\downarrow}^{\uparrow} \; t \left ( \hat{c}^{\dagger}_{j+1,\sigma} \;\hat{c}_{j,\sigma} + \hat{c}^{\dagger}_{j,\sigma} \;\hat{c}_{j+1,\sigma} \right ) + \sum_{j=1}^n \; U \; \hat{n}_{j,\uparrow}\; \hat{n}_{j,\downarrow}$$
where each $\hat{c}^{\dagger}_{j,\sigma}$ ($\hat{c}_{j,\sigma}$) is the fermionic creation (annihilation) operator of site $j$ and spin $\sigma$, $\hat{n}_{j,\sigma}=\hat{c}^{\dagger}_{j,\sigma} \; \hat{c}_{j,\sigma}$, $U$ is the Hubbard on-site repulsion between two electrons and $t$ is related to the kinetic energy of an electron hopping from one site to an adjacent site. A schematic illustration of the structure is shown in the main manuscript in Figure~4g. Approximate values of the two parameters were obtained from the band structure shown in Figure~4f of the main manuscript. The energy spectrum and eigenstates of the system were obtained through exact diagonalization.

\begin{figure}
	 \includegraphics{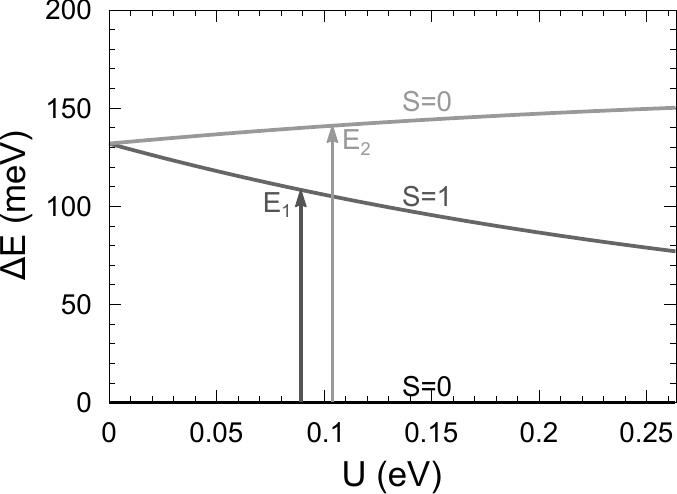}
	\caption{\label{SFig_EvsU}
	Excitation energy $\Delta E$ obtained from the energy spectrum of the Fermi-Hubbard model as a function of $U$ in the two electron model with $t=\SI{133.9}{meV}$.
	}
\end{figure}

\begin{figure}
	\includegraphics[width=1\columnwidth]{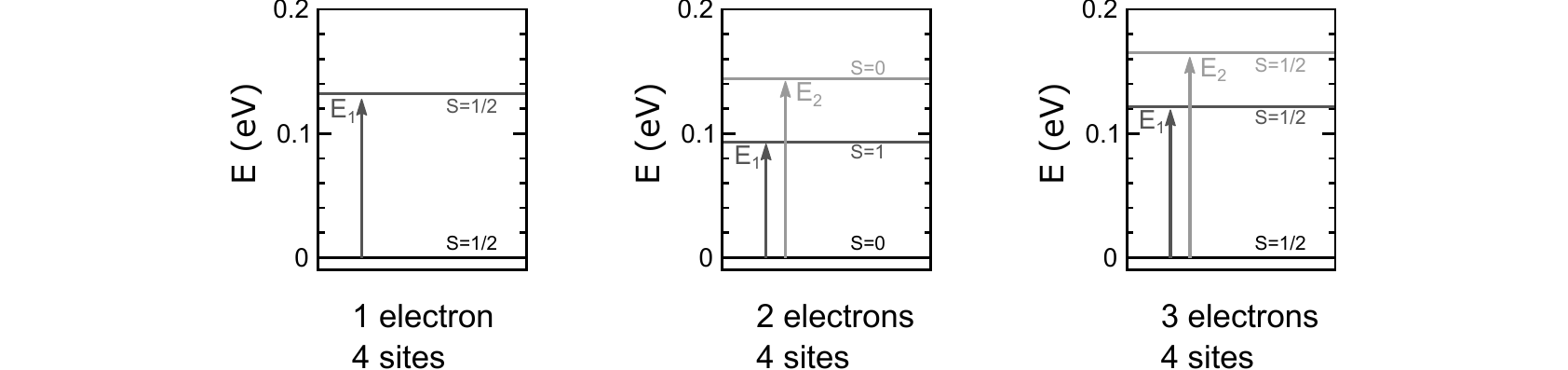}
	\caption{\label{SFig6}
	Excitation spectrum obtained from the four-site model Hamiltonian with one, two and three electrons (left to right) for $U=\SI{155.9}{meV}$ and $t=\SI{131.9}{meV}$. For an odd number of electrons, there are only doublet-doublet excitations in this energy range. For an even number of electrons, there are both singlet-triplet and singlet-singlet excitations. Note that the excitation energy increases when going from two to three electrons.
    }
\end{figure}

To analyze the stability of the Fermi-Hubbard model with respect to changes of $U$, we have considered $U$ values ranging from $0$ to $2t$. The results are presented in Figure~\ref{SFig_EvsU}. Within this energy range, no qualitative change of behaviour is observed. The energy difference between the ground state and the two excited states remains in the order of $t$, indicating that this is the relevant energy scale for the experimentally observed excitations.
Furthermore, the energy scale of the first excited states of the model is robust against changes in the number of electrons. Excitation energies were calculated for one, two and three electrons in the model (Figure~\ref{SFig6}). We find that the excitation energies match reasonably well with the experimental data in all cases.

\FloatBarrier
\clearpage
\section{DFT Calculations}
\subsection{Magnetic States in Finite Ribbons}
\FloatBarrier

\begin{figure}[h!]
	 \includegraphics{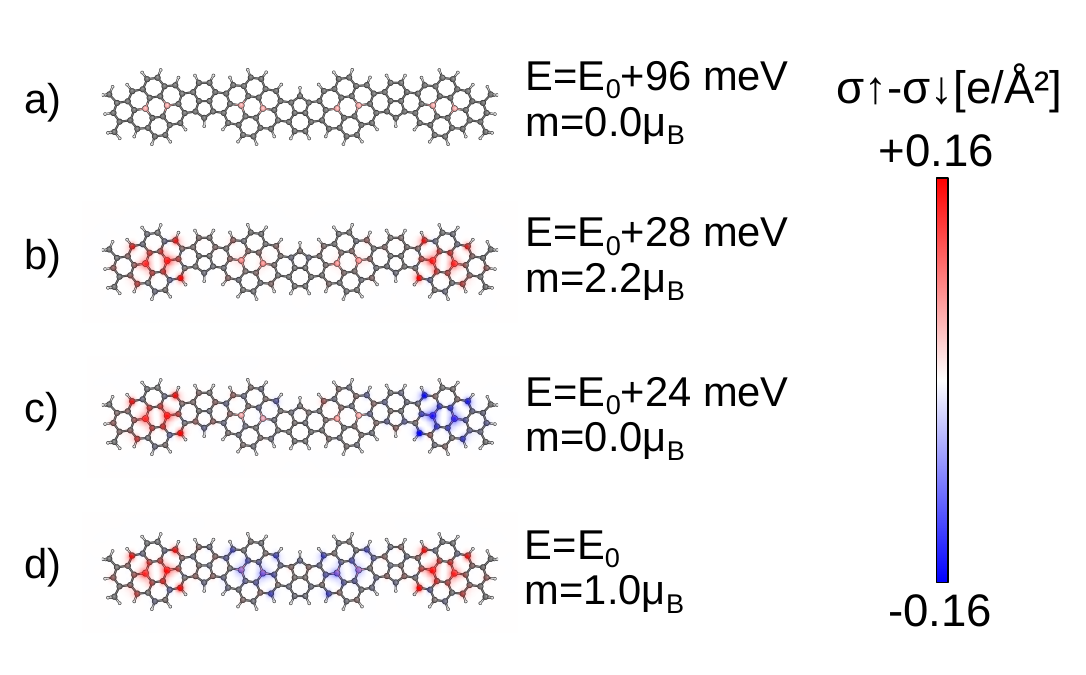}
	\caption{\label{SFig5} 
	Spin polarization density for the finite \pgnr\ converged for different spin configurations.
	(a) Non-spin-polarized solution.
    (b) The spin in all 2B-units aligns. There is a stronger spin-polarization present at the termini.
    (c) The spin in the termini anti-aligns, while there is hardly any spin-polarization localized around the two inner 2B-units.
    (d) Ground state of the system. The spins localized in the termini align with each other and anti-align with the spin around the inner 2B-units. The spin-polarization on the two inner 2B-units does not fully compensate the magnetic moments localized at the termini.
    }
\end{figure}

\begin{figure}[h!]
	\includegraphics[scale=0.6]{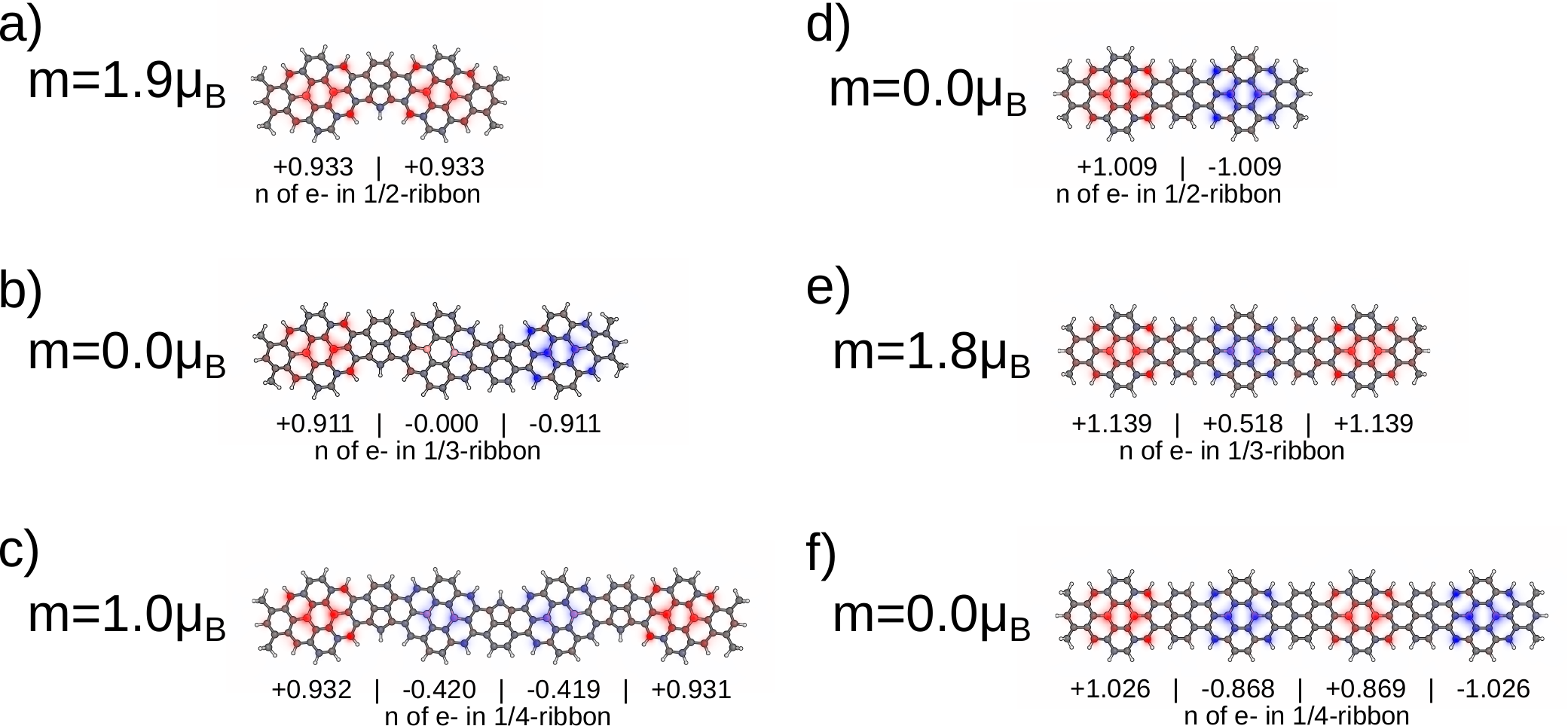}
	\caption{\label{SFig7}
	Spin polarization densities of the energetic ground state of each finite \pgnr\ and \gnr.
	(a),(b),(c) for finite \pgnr\ of 2, 3 and 4 2B-units, respectively.
	(d),(e),(f) for finite \gnr\ of 2, 3 and 4 2B-units, respectively.
    }
\end{figure}

\FloatBarrier
\clearpage
\subsection{Magnetic Ground State of the Periodic \pgnr}
\FloatBarrier

\begin{figure}[h!]
	 \includegraphics{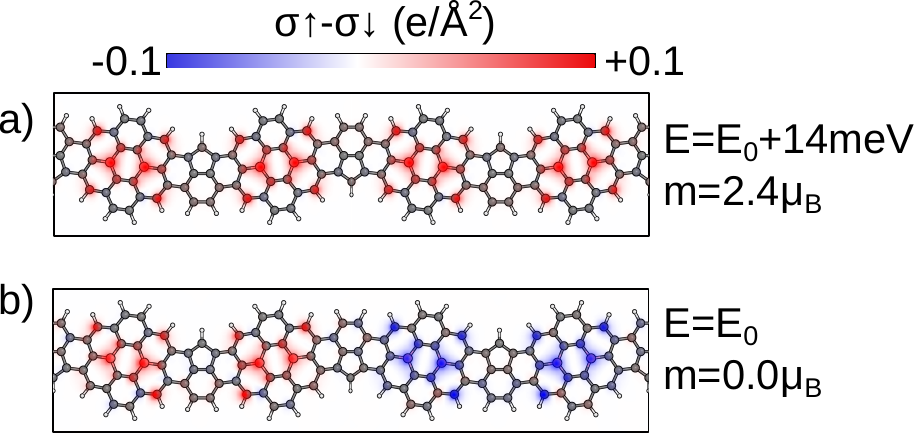}
	\caption{\label{SFig4}
	Spin polarization density for the periodic \pgnr\ using a doubled supercell.
	(a) Aligned solution. 
    (b) Anti-aligned solution. The anti-aligned groundstate is $\SI{14}{meV}$ lower in energy.
    }
\end{figure}
\clearpage
\FloatBarrier
\subsection{Continuous Boron Substitution and Boron Bands Origin}
\FloatBarrier

\begin{figure}[h!]
	\includegraphics[width=1\columnwidth]{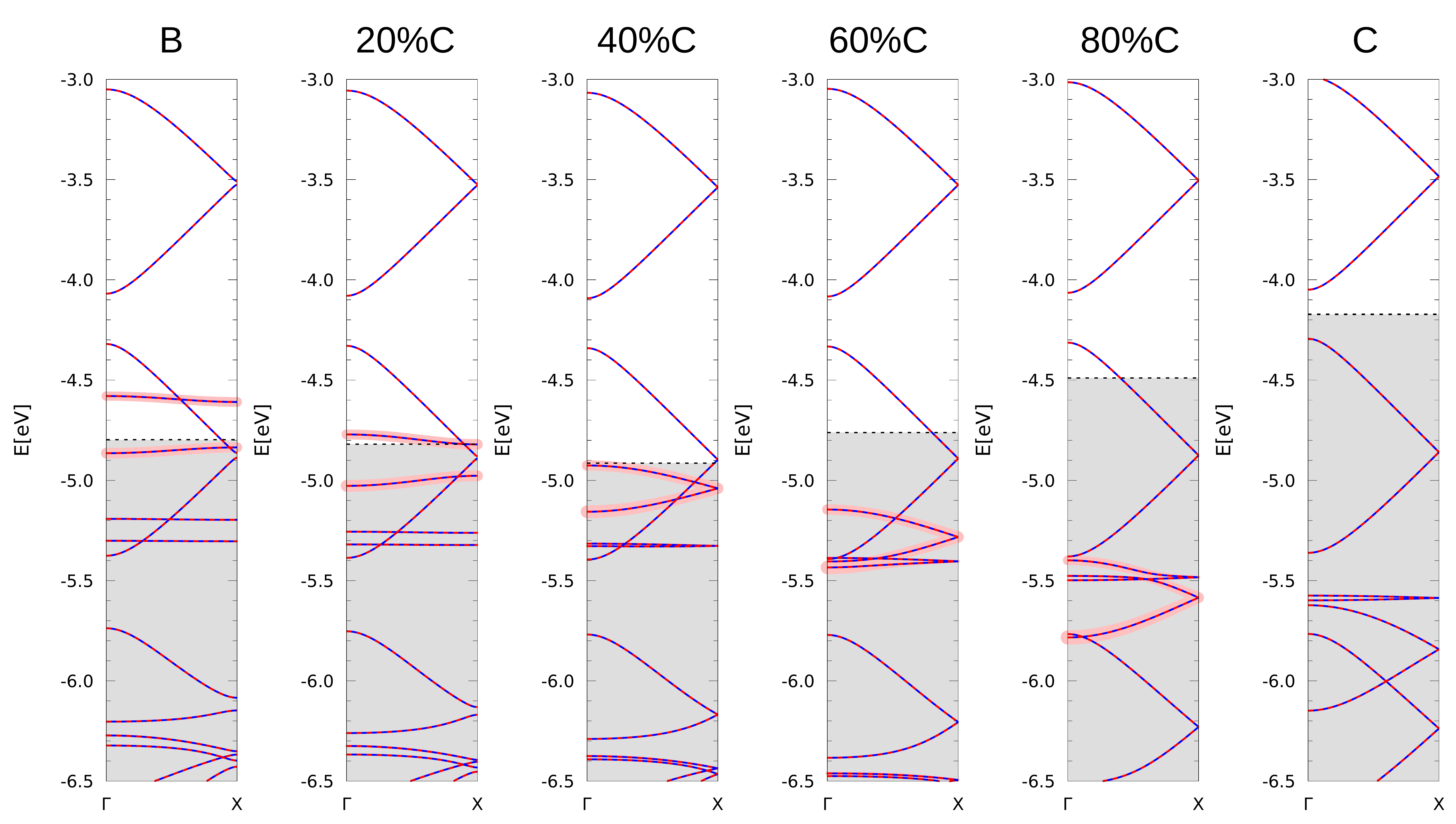}
	\caption{\label{SFig8}
	Electronic band structures of periodic \gnr\ as obtained from virtual crystal calculations with synthetic atoms (created as a mixture of Boron and Carbon) at the substitutional doping sites. Boron character of the bands is indicated by a pink shadow.
    }
\end{figure}

\begin{figure}[h!]
	\includegraphics[width=1\columnwidth]{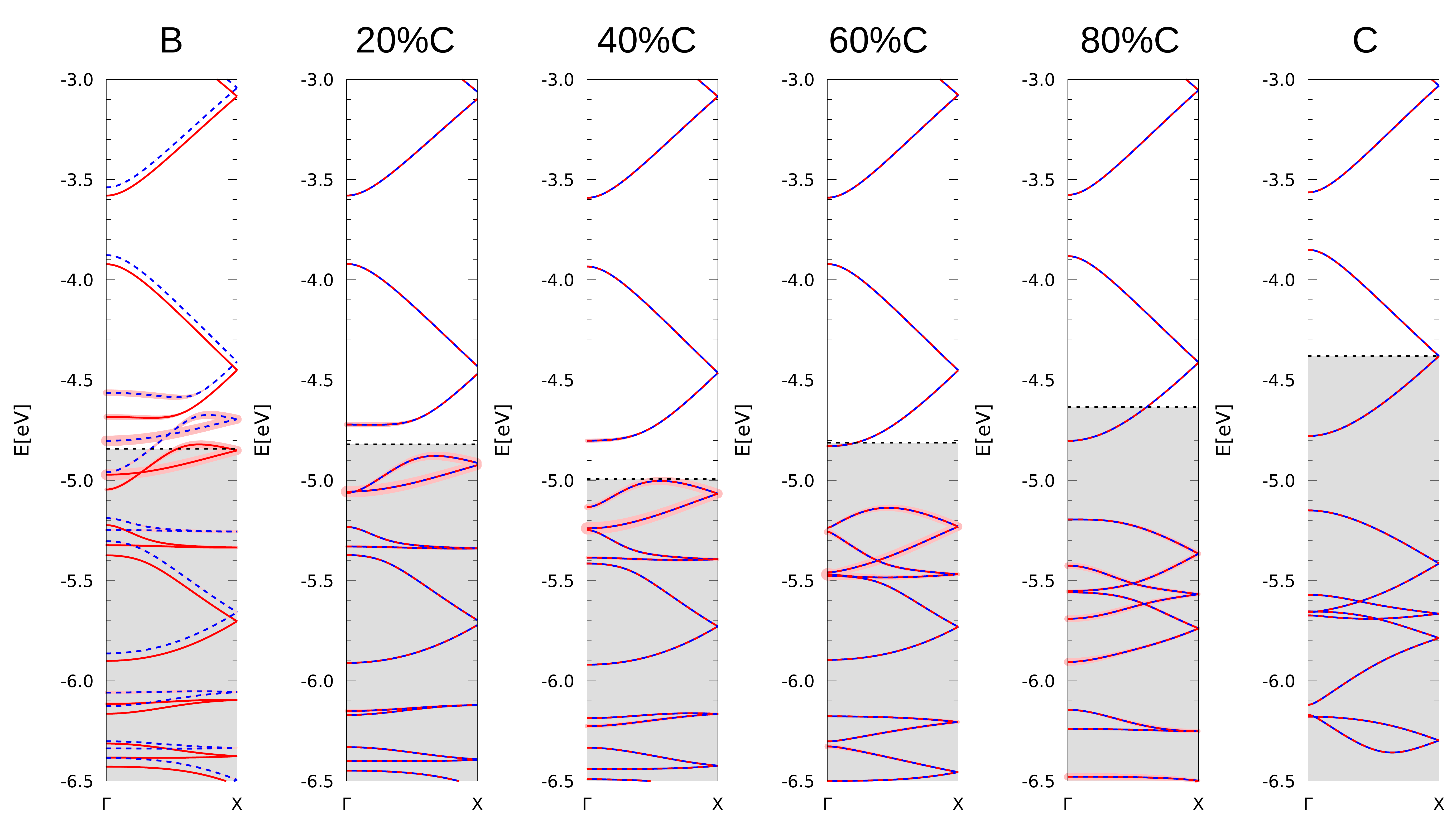}
	\caption{\label{SFig9}
	Electronic band structures of periodic \pgnr\ as obtained from virtual crystal calculations with synthetic atoms (created as a mixture of Boron and Carbon) at the substitutional doping sites. Boron character of the bands is indicated by a pink shadow.
    }
\end{figure}

\clearpage
\FloatBarrier
\subsection{Influence of the Ribbon Width Modulation}
\FloatBarrier

\begin{figure}[h!]
	\includegraphics[scale=0.6]{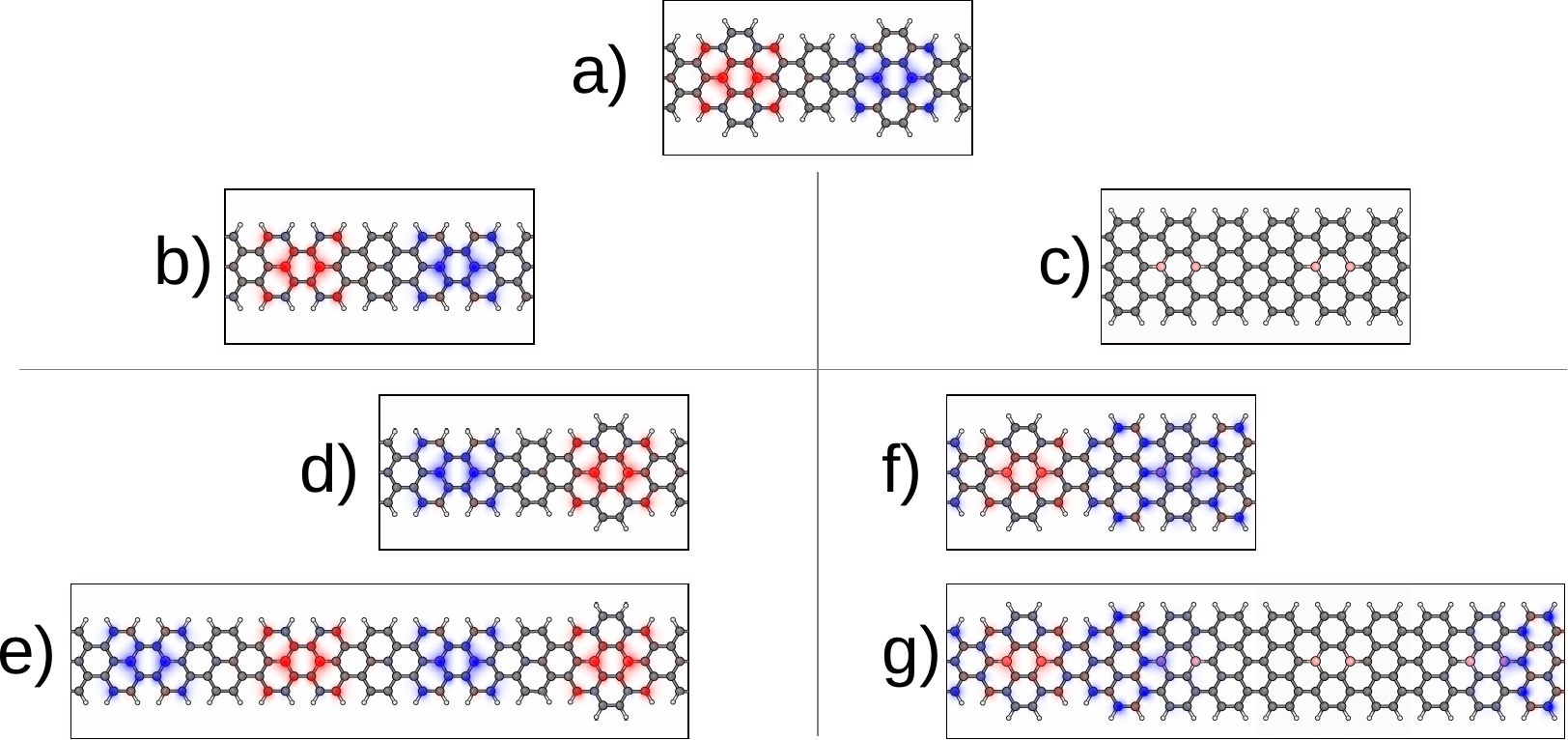}
	\caption{\label{SFig10}
	Spin polarization densities corresponding to different periodic systems with the same distance between substitutional doping sites.
	(a) \gnr, (b) 2B-5-aGNR and (c) 2B-7-aGNR. It is worth pointing out that 2B-5-aGNR exhibits a local magnetization very similar to the one featured in \gnr\ whereas for the 2B-7-aGNR there is no local magnetization.
	Panels (d),(e),(f) and (g) show mixed systems, the combination of \gnr\ with the two other GNRs.
	(d) presents a system combining each one 2B-unit of a 2B-5-aGNR and a \gnr . It presents a magnetization similar to both of their parent systems.
	(e) presents a system combining three 2B-units of a 2B-5-aGNR with one of a \gnr.
	(f) presents a system combining one 2B-unit of a \gnr\ followed by one of a 2B-7-aGNR. The carbon atoms located in the 7-aGNR-segments contribute to the magnetization.
	(g) presents a system combining one 2B-unit of a \gnr\ followed by three of a 2B-7-aGNR. Local magnetization goes to zero for the 2B-7-aGNR unit located the furthest away from the \gnr.
    }
\end{figure}

\clearpage
\FloatBarrier
\subsection{Complete Set of Bands of the \gnr\ Near the Fermi Level}
\FloatBarrier
\begin{figure}[h!]
	\includegraphics[scale=0.8]{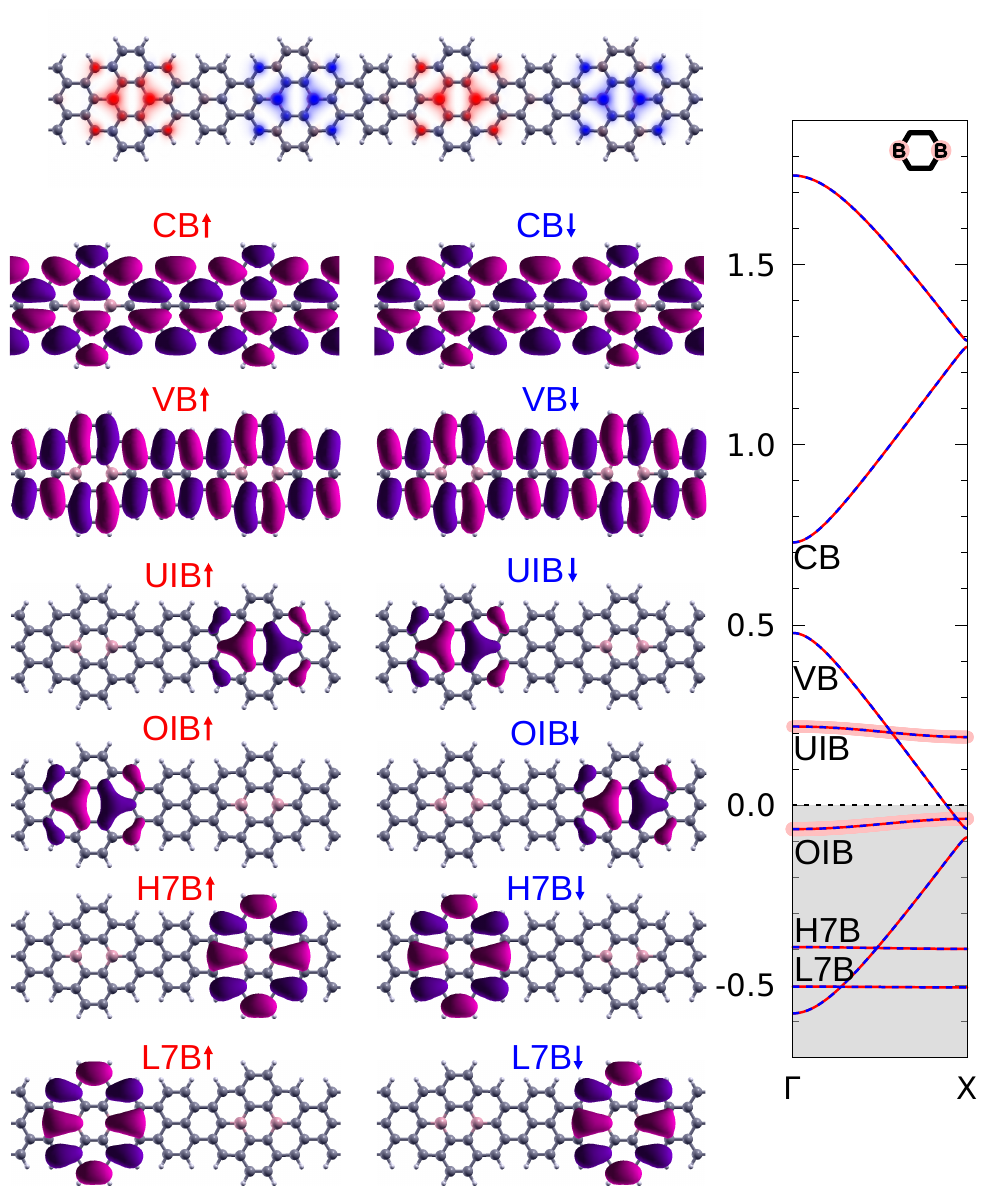}
	\caption{\label{SFigwfcs}
	DFT calculated wavefunctions at $\Gamma$ for the bands near the Fermi level of the \gnr. The wavefunctions for the most dispersive bands (CB and VB) are identical for spin up and down cases. The weakly dispersive bands (UIB, OIB, H7B and L7B) exhibit different wavefunctions depending on the spin.
    }
\end{figure}

\clearpage
\FloatBarrier
\section{Bibliography}

\providecommand{\latin}[1]{#1}
\makeatletter
\providecommand{\doi}
  {\begingroup\let\do\@makeother\dospecials
  \catcode`\{=1 \catcode`\}=2 \doi@aux}
\providecommand{\doi@aux}[1]{\endgroup\texttt{#1}}
\makeatother
\providecommand*\mcitethebibliography{\thebibliography}
\csname @ifundefined\endcsname{endmcitethebibliography}
  {\let\endmcitethebibliography\endthebibliography}{}